\documentclass[pre,preprint,amsmath,amssymb,floatfix]{revtex4}

\usepackage{latexsym,amssymb,amsmath}
\usepackage{graphics}
\usepackage{graphicx}
\usepackage{epsfig}
\begin{document}

\title{Disorder and non-conservation in a driven diffusive system}

\author{M. R. Evans$^1$, T. Hanney$^1$ and Y. Kafri$^2$\\
$^1$ School of Physics, University of Edinburgh,
Mayfield Road, Edinburgh, EH9 3JZ, United Kingdom\\
$^2$ Department
of Physics, Harvard University, Cambridge, MA 02138}

\begin{abstract}
We consider a disordered asymmetric exclusion process in which
randomly chosen sites do not conserve particle number. The model
is motivated by features of many interacting molecular motors such
as RNA polymerases. We solve the steady state exactly in the two
limits of infinite and vanishing non-conserving rates. The first
limit is used as an approximation to large but finite rates and
allows the study of Griffiths singularities in a nonequilibrium
steady state despite the absence of any transition in the pure
model. The disorder is also shown to induce a stretched
exponential decay of system density with stretching exponent
$\phi= 2/5$.
\end{abstract}

\maketitle

\section{Introduction} \label{intro}

Driven diffusive systems serve as simple models for collective
phenomena ranging from traffic flow to molecular motors. Moreover,
they provide tractable examples of systems far from thermal
equilibrium. Studies of one-dimensional driven diffusive systems have
shown that many interesting phenomena, which are typically not
observed in one dimensional systems in thermal equilibrium,
exist. Prominent examples are boundary induced phase transitions and
spontaneous symmetry breaking, for reviews see
e.g. \cite{Mukamel,Brazil,Schutz}.

Most studies have considered systems in which the dynamics are the
same everywhere in the system or systems where the dynamics are
modified only at the boundaries. However, when trying to relate these
systems to many interacting molecular motors, the effects of
non-conservation and disorder (i.e. spatial heterogeneity in the
dynamics) can not be ignored in many cases.

Indeed there have been some studies on the effects of disorder on
driven diffusive systems. For example, the effect of assigning a
disordered quenched rate to each particle was studied in
\cite{KF,E,BJ,JB,TB,K} on a ring geometry. Exact solutions show that
at high enough densities a macroscopic number of particles jam behind
the slowest particle in the system. The phase transition between the
jammed and non-jammed phase is similar to a Bose-Einstein
condensation. Work has also been done on an asymmetric exclusion
process on a ring where the quenched hopping rates between neighboring
sites are drawn at random \cite{TB,K,KP}. For molecular motors moving
along a disordered substrate this seems to be the relevant scenario
\cite{Harms,Kafri}. It was argued, based on numerics and mean-field
solutions, that at high densities the system phase separates into a
region of high density coexisting with a low density region. Finally,
the combined effect of random hopping rates and open boundary
conditions was considered in \cite{ED,HS}.  In \cite{ED} it was argued
using numerics that the location of phase transition lines may be
sample dependent. In \cite{HS} mean-field arguments and numerics
indicate the existence of shifts in phase boundaries which, by analogy
with equilibrium systems, are expected to be accompanied by emergent
Griffiths regions. A review of the effects of disorder in exclusion
models has been given in \cite{S}.

In this paper we consider another type of disorder. We study an
asymmetric exclusion process (ASEP) where {\em non-conserving} sites
are chosen at random along the lattice. At these sites particles may
attach and detach with specified rates which may also be drawn at
random. Thus there are two components to the disorder.  A feature of
this disorder is that it allows a detailed account of the way in which
Griffiths singularities can arise in nonequilibrium steady states.  In
equilibrium the mechanism leading to Griffiths singularities is well
understood: the disorder, e.g. dilution, breaks the system into pure
regions and large pure regions may give rise to the exponentially
suppressed Griffiths singularities. In the present the case, in the
limit of high attachment and detachment rates, the non-conserving
sites break the system into driven conserving domains.

Non-conservation of particles in driven systems without disorder has
previously been considered in the context of molecular motors. The
idea is that molecular motors move in a preferred direction along a
filament and are able to attach and detach from the filament. In the
works so far all sites are non-conserving
\cite{PFF,EJS,LKN,PRWKS}. The motivation for the model we study here
comes from the fact that some molecular motors only attach and detach
at certain sites.

More specifically, we give a very simplified description of
many interacting RNA polymerase (RNAp) motors acting on a prokaryotic
DNA in vitro. Prokaryotic RNA polymerase can initiate without
regulatory proteins. Namely, RNAp left in a solution with DNA can
produce RNA even if the specific protein which regulates its action
(for example, by enhancing or reducing the initiation rates) is
present. They can enter and leave the DNA in order to transcribe RNA
molecules at specific sites, referred to as promotor and termination
sites respectively. In the language of the lattice model we consider a
binding of a RNAp to a promotor corresponds to a particle entering the
system. The unbinding at the termination site corresponds to a
particle leaving the system. In the absence of regulatory proteins the
rate of entering the DNA depends on the details of the promotor
sites. In such systems the RNA polymerase do not usually move from one
gene to another. In the lattice model this would correspond to
particles not moving from one stretch of conserving sites to a
neighboring one i.e. the limit in which the detachment rates are large
at the non-conserving sites. We comment that in principle RNA
polymerase may move in different directions along the DNA when
transcribing different messenger RNAs, corresponding to particles
moving in different directions along different conserving
stretches. However, in the limit when the detachment rate is large
this will not influence most of the results described in the paper. Of
course, our assumption of randomly distributed lengths of genes (or
conserving segments) is not expected to hold. However, the model
provides a starting point for analyzing more realistic situations.

The paper is organised as follows. In Section \ref{model}, we define
the model and discuss two limits which are exactly soluble.  In
Section \ref{GS} we show that the disorder induces Griffiths like
singularities as the rates for entering and leaving the lattice are
changed. More significantly, it is shown in Section \ref{dynamics}
that the presence of the non-conserving sites leads to anomalous
relaxation of the system toward the steady state.  Specifically, we
argue that decay of measurable quantities decay as a function of time,
$t$, as a stretched exponential $\exp(-ct^\phi)$, where $c$ is a
non-universal constant and $\phi=2/5$. The results are verified
numerically. We conclude in Section \ref{CONC}.

\section{Model} \label{model}

The model we consider is a disordered generalisation of the ASEP. The
pure ASEP is defined on a one-dimensional lattice containing $L$ sites
and with periodic boundaries. The lattice is occupied by particles
subject to an exclusion interaction, which prohibits multiple
occupancy of any site. These particles hop with rate one to the
nearest neighbour site to the right, provided it is empty, and so the
total particle number $N$ is conserved.  We introduce non-conservation
into this model by allowing, at certain sites (which we will call
`disorder' sites), processes which do not conserve the total particle
number $N$. Hence each site $l$ ($l=1,\ldots,L$) in the pure model
remains a pure site with probability $p$, or becomes a disorder site
with probability $(1-p)$. Now, at the disorder sites, labelled by
$j=1,\ldots,P$, particles attach with rate $c_j$ or detach with rate
$a_j$. In general, we wish to consider heterogeneous rates for the
non-conserving processes.

To study the model we first consider limits which can be solved
exactly. Later, using numerics, we argue that the results are
generic. Exact solubility arises when the steady state densities at
the disorder sites are determined solely by the attachment and
detachment processes.  In these cases, since the system is composed of
conserving domains of the chain in contact with disorder sites at the
boundaries of each domain, the steady state can be written as a
product of boundary driven ASEPs in which the densities of the
boundary reservoirs are given by the disorder site densities $\rho_j$.

Before turning to the disordered ASEP under consideration we recap
some facts, which will be useful later, about the boundary driven
ASEP. For the boundary driven ASEP, in which particles are injected at
the left-hand boundary site with rate $\alpha$ (provided it is vacant)
and removed from the right-hand boundary site with rate $\beta$, exact
steady state weights for particle configurations can be obtained using
a matrix product ansatz \cite{DEHP}. In this ansatz, particle
configurations are represented as a product of matrices $X_1 \cdots
X_L$ where $X_l = D$ ($E$) if site $l$ is occupied (vacant). The
steady state weight of a configuration is given by $\langle \alpha
\vert X_1 \cdots X_L \vert \beta \rangle$ provided the matrices $D$
and $E$ and the vectors $\langle \alpha \vert$ and $\vert \beta
\rangle$ satisfy the relations
\begin{equation}
DE=D+E \equiv C\;, \quad \alpha \langle \alpha \vert E =
\langle \alpha \vert \;\quad {\rm and} \;\quad \beta D \vert \beta \rangle =
\vert \beta \rangle\;.
\end{equation}
From these relations exact expressions for the normalisation $\langle
\alpha \vert C^L \vert \beta \rangle$ can be derived which show that
the model undergoes a second order phase transition: when both
$\alpha$ and $\beta \geq 1/2$ the system is in a maximum current
phase, otherwise it is in one of two low current phases. The phase
transition between the low current phases is first order. The rates
$\alpha$ and $1-\beta$ represent the densities of particles in
reservoirs connected to the boundary sites.

Next, we use the known results for the boundary driven ASEP to
study the disordered case. As stated above, there are limits where
the steady state weight of the disordered model factorizes into a
product over boundary driven ASEP weights. The two
exactly soluble limits are:

\begin{itemize}
\item $c_j$, $a_j \to \infty$, with $c_j/a_j$ fixed.

In this limit, each  disorder site $j$ acquires a density $\rho_j$ determined
solely by  $c_j$ and $a_j$ which
obeys the equation of motion,
\begin{equation}
\frac{\partial \rho_j}{\partial t} = c_j (1-\rho_j) + a_j \rho_j\;.
\end{equation}
Therefore in the steady state,
\begin{equation}
\rho_j = \frac{c_j}{c_j+a_j}\;.
\label{rho}
\end{equation}
If we define $n_j$ to be the number of sites between disorder
sites $j$ and $j+1$ (i.e. the length of the $j$-th conserving domain),
then the normalisation $Z_L(\{n_j\})$ (which is the sum over the
steady state weights of all particle configurations on sites excluding
the disorder sites), for a given configuration of the disorder sites
$\{n_j\}=n_1,\ldots,n_P$, factorises into a product over
normalisations for the boundary driven ASEP:
\begin{equation} \label{infty}
Z_L(\{n_j\}) = \prod_{j=1}^P \langle \rho_j \vert C^{n_j} \vert 1-\rho_{j+1}
\rangle
\end{equation}
and where $\rho_{P+1} = \rho_1$.
This is the relevant limit for the model molecular motors discussed
in the introduction.

\item $c_j$, $a_j \to 0$, with $c_j/a_j$ fixed.

In this limit the time between each attachment/detachment event
tends to infinity. Therefore, after each event the system reaches
a homogeneous steady state of the pure ASEP  with periodic
boundaries. Thus the system density $\rho = N/L$,  satisfies the
equation of motion
\begin{equation}
\frac{\partial \rho}{\partial t} = \left[ \sum_{j=1}^P c_j \right]
(1-\rho) +\left[ \sum_{j=1}^P a_j \right] \rho\;.
\end{equation}
Therefore in the steady state the system density is given by
\begin{equation}
\rho = \frac{\sum_{j=1}^P c_j}{\sum_{j=1}^P (c_j+a_j)}\;.
\end{equation}
Because the steady state is homogeneous, all sites, including
disorder sites, have the same steady state density $\rho$.
Moreover, the steady factorises and there are no correlations
between sites. Therefore, one can still write the normalisation
in a form similar to (\ref{infty}):
\begin{equation} \label{zero}
Z_L(\{n_j\}) = \prod_{j=1}^P \langle \rho \vert C^{n_j} \vert 1-\rho
\rangle\;.
\end{equation}
This is because in this case $D$ and $E$ are given by
the  scalars
$1/(1-\rho)$ and $1/\rho$.
\end{itemize}

In the following we will consider the first limit, $c_j$, $a_j \to
\infty$. We use the factorised form (\ref{infty}) with $\rho_j$ given
by (\ref{rho}) as an approximation for the case where $c_j$, $a_j$ are
large but finite which is relevant for the model of molecular motors.
This approximation has a mean-field character, in the sense that
correlations are factorised about the disorder sites, however all
correlations within conserving domains are retained.

\section{Griffiths singularities} \label{GS}

We can exploit known properties of the normalisation of the boundary
driven ASEP to demonstrate the existence of Griffiths-type
singularities in the disordered ASEP \cite{G,M}. As an illustrative
example, we consider binary disorder at disorder sites, such that
\begin{displaymath}
\rho_j = \frac{c_j}{a_j+c_j} = \left\{ \begin{array}{ll} u & \textrm{
    with probability $q$,} \\ v & \textrm{with probability $1-q$.}
    \end{array} \right.
\end{displaymath}
This is the simplest choice of disorder for which Griffiths
singularities occur. Generalizations to more complicated
situations are straightforward.

Using (\ref{infty}) the steady state normalization satisfies
\begin{equation}
{\rm ln}Z_L = \sum_{j=1}^P W_{n_j} (\rho_j, 1\!-\!\rho_{j+1}) \;,
\label{lnZ}
\end{equation}
where $W_n (\rho_j, 1-\rho_{j+1}) = {\rm ln} \langle \rho_j \vert C^{n_j}
\vert 1-\rho_{j+1} \rangle$. In order to perform the disorder average, we
write (\ref{lnZ}) in the form
\begin{eqnarray}
{\rm ln}Z_L = \sum_{n=0}^\infty  & \left[ \nu_{u,1-v}(n) \, W_n
  (u,1\!-\!v) + \nu_{u,1-u}(n) \, W_n (u,1\!-\!u) \right. \nonumber \\
  & \left. + \nu_{v,1-v}(n) \, W_n (v,1\!-\!v) + \nu_{v,1-u}(n) \, W_n
  (v,1\!-\!u) \right]
\end{eqnarray}
where $\nu_{\alpha,\beta}(n)$ is the number of conserving domains of
size $n$ bounded by disorder sites at densities $\alpha$ and $1-\beta$.

We can average over the configurations of the $\nu_{\alpha,\beta}(n)$
by calculating the expectation values $\langle
\nu_{\alpha,\beta}(n) \rangle$ in the thermodynamic limit (the angled
brackets denote a disorder average). This is
achieved by observing that $\lim_{L \to \infty}
L^{-1}\langle \nu_{\alpha,\beta}(n) \rangle$ is just the probability that
a site is part of an $n$-site conserving domain bounded by disorder
sites with densities $\alpha$ and $1-\beta$, hence
\begin{eqnarray} \label{w}
\lim_{L \to \infty} L^{-1} \langle {\rm ln} Z_L \rangle = (1\!-\!p)^2
\sum_{n=0}^\infty p^n &\left\{ q^2
W_n(u,1\!-\!u) + q(1\!-\!q)
    [W_n(u,1\!-\!v)  \right.\nonumber \\ &\left. + W_n(v,1\!-\!u)] + (1\!-\!q)^2
    W_n(v,1\!-\!v) \right\}
\end{eqnarray}

The form of equation (\ref{w}) is typical of systems which exhibit
Griffiths singularities. In equilibrium, these singularities are
usually inferred from the properties of the Yang-Lee zeros of the
partition function --- we can use the known properties of the Yang-Lee
zeros of the the analogous quantity, the normalisation \cite{BE}, to
show how Griffiths singularities arise in the disordered
nonequilibrium model: For fixed $u \geq 1/2$ say, in the complex
$v$-plane and for $n$ arbitrarily large, the zeros of $\langle u \vert
C^n \vert 1-v \rangle$ accumulate arbitrarily close to the point
$v=1/2$ on the real axis. Therefore there exists a singularity in
$W_n(u,1-v)$ arbitrarily close to the point $v=1/2$ which is
exponentially suppressed (by a factor $p^n$). Thus, such a
Griffiths-type singularity follows whenever $u$ and $v$ are such that
at least one of the $W_n(\alpha,\beta)$ in (\ref{w}) lies on the phase
boundary of the ASEP i.e. whenever $u$ and/or $v=1/2$.

One can go further and consider disorder in the $c_j$'s and $a_j$'s
explicitly. For instance, if both $c_j$ and $a_j$ are drawn from
binary distributions, then the densities at the disorder sites can
assume one of four possible values, each with a different probability
in general. However Griffiths singularities still arise whenever any
one of these values for the density is 1/2, as before.

It is also straightforward to use standard arguments from the study of
dilute systems to show that the correlation length remains finite at
the Griffiths singularity, as is the case in equilibrium systems. Note
that Griffiths singularities emerge in this non-equilibrium system
despite the absence of a transition in the pure model.  This is in
sharp contrast to equilibrium systems.

\section{Dynamics: stretched exponential decay of the density to its
  steady state value} \label{dynamics}

In the pure boundary driven ASEP, whenever the system is in or at the
boundary of the maximum current phase, the system density decays with
time to its stationary value as an exponential with a decay constant
that depends on system size $L$ as $L^{z}$, where $z=3/2$ is the
dynamic exponent\cite{BW,NAS}.  In the low current phase when the
boundary injection and ejection rates are equal, a shock exists in the
steady state, and the dynamic exponent $z=2$. Otherwise, in the low
current phases the relaxation time is finite \cite{BW} and does not
depend on $L$ 
Hence, in the disordered model, whenever
contributions to the normalisation (\ref{w}) are in the maximum
current phase, the decay of the system density will be determined by
the relaxation of these conserving domains.

For the following analysis, it is sufficient to consider disorder only
in the location of the disorder sites: we consider homogeneous
attachment and detachment rates i.e. $c_j=c$ and $a_j=a$.  In the case
where $c=a$, (\ref{rho}) gives $\rho = 1/2$ at which point conserving
domains are on the boundary of the maximum current phase; otherwise
the conserving domains are in the low current phases.  Thus only when
$c=a$ do we expect the decay constant associated with a conserving
domain to depend on its size\cite{NAS}.

Therefore for $c=a$, in a conserving domain of length $n$, we assume
that the particle density $\rho_n(t)$ decays to its steady state value
as
\begin{equation}
\delta\rho_n(t) \equiv \rho_n(t) - 1/2  \sim e^{-\Delta_n t},
\end{equation}
where $\Delta_n = \Delta_0 n^{-3/2}$.
In the disordered case, we need to sum
over configurations of the disorder sites. This is achieved in the
same way as in the previous section so, in the thermodynamic limit, the
decay of the system density  $\rho(t)$ becomes
\begin{equation}
 \delta\rho(t) \sim \sum_{n=0}^{\infty} p^n  e^{-\Delta_n t}\;,
\end{equation}
where we retain only the $n$-dependence, as is sufficient to determine
the dominant contribution to the form of the relaxation. If we
convert the sum into an integral and consider late times so that the
integral can be evaluated at the saddle point, we obtain
\begin{equation} \label{decay}
\delta\rho(t) \sim {\rm exp} (-c t^\phi),
\end{equation}
where $c$ is a constant and $\phi = (1+z)^{-1} = 2/5$.

Equation (\ref{decay}) predicts the decay of the density up to some
prefactor, power-law in $t$, with an exponent peculiar to the decay of
the density. The stretching exponent
$\phi$ should be universal however, in the sense that other
correlation functions 
e.g. the current, should reach their
stationary values with the same stretched exponential decay.
This result should be valid for more general types of disorder
whenever one has conserving domains in the maximal current phase.

\begin{figure}[!ht]
\begin{minipage}[c]{0.48 \linewidth}
\includegraphics[width=0.95\textwidth]{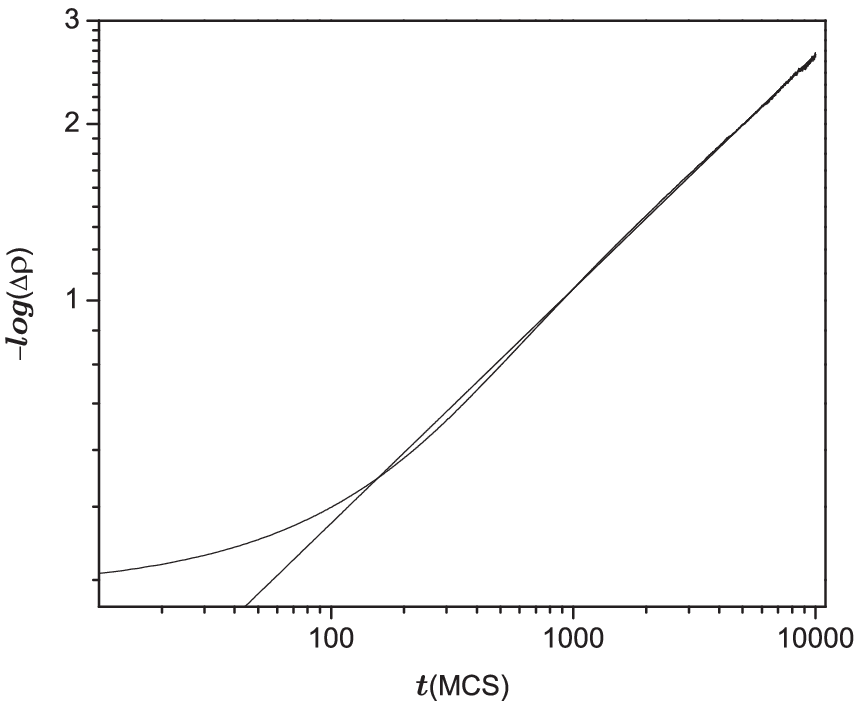}
\caption{Log-log plot of the  decay of the density
 with time  for $a_j = c_j = 10$; the initial condition was an empty lattice.}
\label{fig:decay10}
\end{minipage}
\hfill
\begin{minipage}[c]{0.48 \linewidth}
\includegraphics[width=0.95\textwidth]{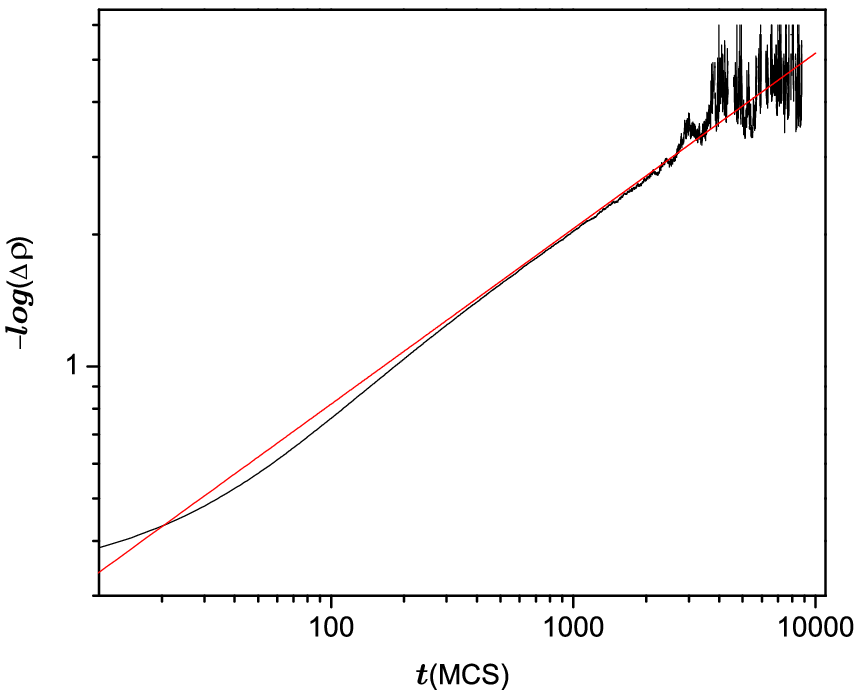}
\caption{Log-log plot of the  decay of the density
 with time  for $a_j = c_j = 1$; the initial condition was an empty lattice.}
\label{fig:decay1}
\end{minipage}
\end{figure}

In figures \ref{fig:decay10} and \ref{fig:decay1} we show the results
of simulations. The simulations were run on systems of 10000 sites
with periodic boundary conditions and averaged over 1000 histories of
the dynamics, starting from an empty lattice and with the same
realisation of disorder.  The decay of the averaged system density
$\delta \rho(t)$, is shown in Figure \ref{fig:decay10} for $a_j = c_j
= 10$, and in Figure \ref{fig:decay1} for $a_j = c_j = 1$.  In both
cases the straight line $t^{2/5}$ is given for reference. Figure
\ref{fig:decay10} shows very good agreement with the predicted
stretched exponential decay, and even in figure \ref{fig:decay1},
where $c_j$ and $a_j$ are not large, the agreement is still quite
good.  Thus it appears that our result for the stretching exponent
holds for finite rates, although its derivation is only exact in the
limit of infinite attachment and detachment rates.

\section{Conclusion} \label{CONC}
In this work we have studied an ASEP with disorder sites where
particles are not conserved. This may provide a basis for a more
realistic model for interacting molecular motors such as RNAp.  We
have used an approximation, exact in the limit of infinite
non-conserving rates, the underlying assumption behind which is
that the system factorises into conserving domains. According to
the ratios of the attachment and detachment rates these domains
assume different phases of the ASEP with open boundaries.

Within this approximation we can explicitly identify Griffiths
singularities.  These arise when there are large conserving domains, on
the boundary of the maximal current phase. An interesting feature is
the prediction of a Griffiths singularity despite the absence of a
transition in the pure system.

More generally one might ask under what conditions do Griffiths
singularities arise in nonequilibrium steady states.  In equilibrium
systems Griffiths singularities are understood in terms of Yang-Lee
zeros of the partition function.  In nonequilibrium systems one does
not have an energy function, nevertheless one can often identify a
quantity that plays the role of a partition function, for example the
normalisation (\ref{infty}), and recently there has been progress in
understanding the zeros of such quantities \cite{BE}.

When there is a spectrum of maximal-current conserving-domain sizes,
we have demonstrated that correlation functions undergo a stretched
exponential decay with a stretching exponent predicted to be
$\phi=2/5$. Moreover simulations suggest this result holds
for a wide range of attachment and detachment rates.
A related stretched exponential decay has already been
observed for the decay of autocorrelations in a bond diluted symmetric
exclusion process on a ring \cite{GS} (in this case $z=2$ so
$\phi=1/3$).

It would be instructive to develop further the approximation that the
steady state factorises about the disorder sites.  As we saw in
Section \ref{model} this approximation is exact in two limits, and we gave
expressions for the densities at the disorder sites. It would be
interesting to develop a scheme that interpolates between these two
limits. Also of interest would be a better understanding of the
correlations between the conserving domains which may exist away from
the two exact limits and their effect on Griffiths singularities.

{\bf Acknowledgments} The authors thank the Max Planck Institute for
Complex Systems, Dresden, where this work was initiated, for
hospitality. Work by Y.K was supported by the National Science
Foundation through grant DMR-0229243 and grant DMR-0231631 and the
Harvard Materials Research Laboratory via Grant DMR-0213805.  MRE and
TH were supported by EPSRC programme grant GR/S10377/01.

\end{document}